\documentclass[aps,prl,floatfix,twocolumn,footinbib]{revtex4}
\usepackage{epsfig}
\usepackage{amsmath,amssymb}
\begin{document}

\title{Proposal for a digital converter of analog magnetic signals
}

\author{Christian Ertler\footnote{email: christian.ertler@physik.uni-regensburg.de}}
\author{Jaroslav Fabian\footnote{email: jaroslav.fabian@physik.uni-regensburg.de}}
\affiliation{Institute for Theoretical Physics, University of
Regensburg, D-93040 Regensburg, Germany}

\begin{abstract}
A device which converts analog magnetic signals directly into
digital information is proposed. The device concept is based on the
monostable-bistable transition logic element, which
consists of two resonant tunneling diodes (load  and driver)
connected in series and employs the monostable to bistable working
point transition of the circuit. Using a magnetic resonant
tunneling diode as the driver allows to control the resulting working
point of the bistable region by an external magnetic field leading
either to high or low output voltage of the circuit, effectively
realizing what could be called digital magnetoresistance.

\end{abstract}
\maketitle

Since the realization of magnetic III-V semiconductors
\cite{Oh98} substantial progress has been made in pushing the transition
temperature to ever higher values, strongly nurturing the hope for
technically applicable room temperature magnetic semiconductors in
near future. Much has been done to exploit the possibilities
offered by these materials in diverse spintronic device concepts
\cite{ZuFaSa04}. For instance, in magnetic resonant tunneling diodes
(m-RTDs) \cite{SlGoSl03, HaTaAs00,PeChDe02, BrWu98, VuMe03, VoShLe00,
BeBeBo05} or magnetic multiple quantum wells diodes
\cite{SaMaPl01, ErFa06}, the transmission can strongly depend on the
spin orientation of the electrons at the Fermi level, which allows
to use the diodes as spin filters and detectors. The quantum well of RTDs
can be formed either by a ferromagnetic
semiconductor \cite{OiMoKa04, FuWoLi04, Di02} or by dilute magnetic
semiconductors (DMSs)\cite{Fu88}, which exhibit giant $g$ factors.
Magnetic devices provide the interesting opportunity of realizing nonvolatile
reprogrammable logic elements. A nonvolatile ferromagnet/superconductor switch based on
Andreev reflections was proposed in \cite{NaMa02}. 
 
Nonmagnetic RTDs have been used for multiple
valued logic and multiple logic functions with multiple input and/or
output \cite{MaFo:book03}, due to their specific N-shaped
current-voltage (IV) characteristics and their extremely high-speed
potential. In 1993 Maezawa and Mizutani proposed the MOnostable-BIstable 
transition Logic Element (MOBILE) \cite{MaMi93}. The concept was extended to 
multistable elements \cite{Waho95,WaChYa98} and applications
to ultrahigh-speed analog to digital converters
were demonstrated \cite{HaTaWa02}.
The MOBILE device consists of two nonmagnetic RTDs, 
a load and a driver, which are connected in series. In this
paper we demonstrate that MOBILE circuits with m-RTDs can be used as
digital converters of analog magnetic signals by realizing digital
magnetoresistance (DMR): for a continuous change of the magnetic
field through a {\em controlled} threshold the output electrical
characteristics such as voltage exhibit discrete jumps. A nice related effect of the MOBILE 
has been realized by Hanbicki et al. \cite{HaMaCh01}: a nonmagnetic driver RTD is shunted with 
a metallic giant magnetoresistance resistor, allowing for current modulations through the RTD. This
fascinating phenomenon makes DMR-MOBILEs serious candidates for
magnetic reading devices. 

\begin{figure}
 \centerline{\psfig{file=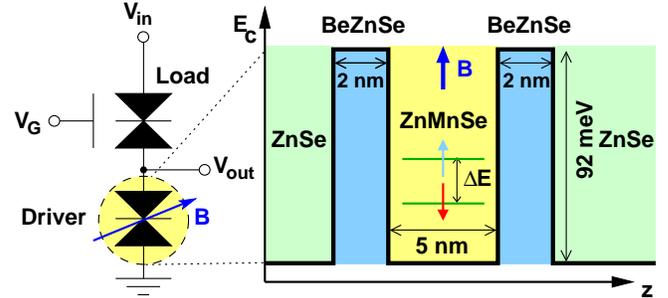,width=1\linewidth}}

\caption{(Color online) Left: The circuit configuration of the
proposed DMR-MOBILE. The load is a conventional RTD, whose peak
current can be modified by an external gate voltage $V_G$. The
driver device consists of a magnetic RTD (here made of a Zn(Be,Mn)Se
material system). The peak current of the driver is controlled by the
magnetic field. Right: The schematic conduction band profile of the
magnetic RTD  used in the numerical
simulations discussed in the text.} \label{fig:circuit}
\end{figure}
%
%

%
%

\begin{figure}
 \centerline{\psfig{file=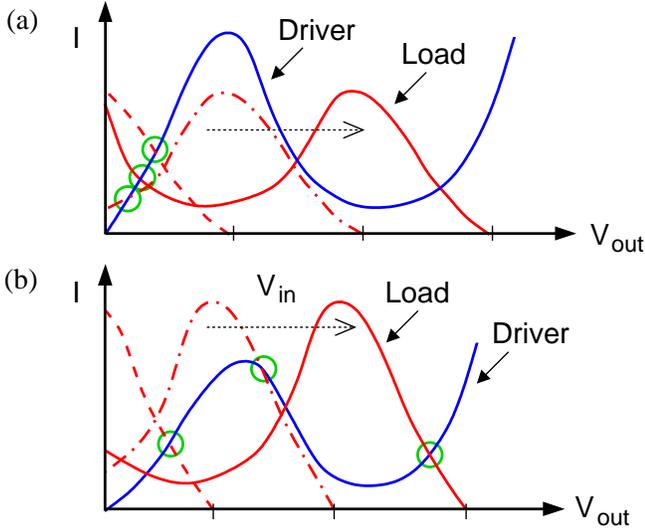,width=1\linewidth}}

\caption{(Color online) Schematic load line diagrams for the case of
a  higher (a) and lower (b) driver peak current compared
to the load peak value. The mirrored load IV-characteristic is
shifted from the left (dashed line) to the right (solid line) when
the input voltage $V_{\mathrm{in}}$ is increased. The dot-dashed
line refers to an intermediate input voltage. The actual (stable)
working point of the circuit is marked by a circle. }
\label{fig:transition}
\end{figure}

Figure \ref{fig:circuit} shows the
schematic circuit diagram for a DMR-MOBILE, in which the driver is
a m-RTD. The basic concept of the operation of a MOBILE is to drive the
circuit by an oscillating input voltage $V_{\mathrm{in}}$ to produce
the transition between the monostable and bistable working points
regime of the circuit \cite{MaMi93}. In the bistable regime there is
a stable DC working point at a low voltage and another at a high
voltage. Which of them is actually realized after the transition by
the circuit is determined by the difference of the peak currents
between the two RTDs. If the load peak current is lower than the
driver peak current the circuit output voltage is low, and vice
versa. The load line diagrams of both cases are schematically shown
in Fig. \ref{fig:transition}. The upper (lower) diagram shows the
case when the peak current of the load is smaller (higher) than that of
the driver device. By increasing the input voltage the mirrored IV
of the load is shifted rightwards. The intersection points with the
driver IV are the candidates for stable DC working points of the
circuit. The number of crossing points changes from one at low input
voltages (monostable regime) up to three at high input voltages
(bistable regime). However, the crossing point in the negative
differential resistance (NDR) region proves to be unstable by the
following criterion. For small fluctuations $\delta V_l$ and $\delta
V_d$ from the equilibrium load and driver voltages $V_l$ and $V_d$
(with the constraint $V_l+V_d = V_{\mathrm{in}}$), a Rayleigh
dissipative potential can be defined as
\begin{equation}
\delta P = \frac{1}{2}[G_l (\delta V_l)^2+G_d (\delta V_d)^2]=
\frac{1}{2} (G_l+G_d)(\delta V_d)^2,
\end{equation}
where $G_{l}$ and $G_d$ denote the differential conductances at the
working point of the load and driver, respectively. For a stable
system, the dissipative potential must be a positive definite form
leading to the stability criterion $G_l+G_d > 0$, which is clearly
violated in the NDR-region. A more detailed  analysis (including the
dynamic behavior of the circuit) of the stability of working points
in RTDs is given in \cite{KiMeEa91, Ch:book64}. As illustrated in
Fig. \ref{fig:transition}(a) the actual working point (marked by a
circle) remains always in the low voltage area if the load peak
current is smaller than the driver peak value. For the opposite
case, displayed in Fig. \ref{fig:transition}(b), the working point
can overcome the driver peak voltage, resulting in a high output
voltage for high input voltages. The difference in the peak currents
can be very small to perform the switching between the low and high
voltage states, as the transition is in some sense analogous to a
second-order phase transition \cite{MaMi93}.

For our proposed device, DMR-MOBILE, we suggest to use a
conventional RTD for the load but to replace the driver by a
m-RTD as illustrated in Fig. \ref{fig:circuit}. This allows,
as we will see below, to change the peak current of the driver by an
external magnetic field. In addition, the peak current of the load
is tunable by an external gate voltage \cite{MaMi93}. The schematic
conduction band profile of the m-RTD is also shown in Fig
\ref{fig:circuit}. The quantum well is considered to be made of the
DMS material ZnMnSe, whereas the
barriers are formed by Be-doped ZnSe. The barrier height is assumed
to be $92$ meV, which is about 23\% of the band gap difference
between ZnBeSe and ZnSe. The active region of the structure is
sandwiched between two $n$-doped ZnSe layers with $n\approx 10^{18}$
cm$^{-3}$ including a 5 nm thick undoped buffer layer. Similar all
II-VI semiconductor m-RTDs have already been experimentally
investigated \cite{SlGoSl03}.

Due to the giant $g$ factor in DMSs an external magnetic field $B$
causes a giant Zeeman energy splitting of the conduction band. The
energy difference of the spin up and down states $\Delta E$ can be
expressed by a modified Brillouin function $B_s$ \cite{BeBeBo05}:
\begin{equation}
\Delta E  = x_{\mathrm{eff}} N_0\alpha s B_s(g_{Mn} s \mu_B B/k_B
T_{\mathrm{eff}}),
\end{equation}
where $x_{\mathrm{eff}}$ is the effective concentration of Mn ions,
$N_0\alpha = 0.26$ eV is the $sp-d$ exchange constant for conduction
electrons, $s= 5/2$ is the Mn spin, $g_{Mn} =2.00$ is the Mn
$g$ factor, $\mu_B$ labels the Bohr magneton, $k_B$ denotes the Boltzmann
constant, and $T_{\mathrm{eff}}$ is an effective temperature. Hence,
the DMS gives rise to a spin dependent potential energy term in the
electron Hamiltonian, $U_\sigma(z) = \sigma\Delta E(z)$, with $z$
denoting the growth direction of the heterostructure and $\sigma =
\pm 1/2 $ or $ (\uparrow,\downarrow)$ labeling the spin quantum
number. Assuming coherent transport through the m-RTD we calculate
the spin dependent current  by numerically solving the single band
effective mass Schr\"odinger equation. Space charge effects are
self-consistently taken into account in our numerical simulations by
calculating the conduction band profile from the Poisson equation.
The spin dependent current density $J_\sigma$ is then calculated by
the Tsu-Esaki formula \cite{TsEs73}:
\begin{eqnarray}
J_\sigma & = &\frac{e m^*k_BT}{4\pi^2\hbar^3}\int_0^\infty \mathrm{d} E_z T_\sigma(E_z) g(E_z),\nonumber\\
g(E_z) & =& \ln\left[\frac{1+\exp(E_f-E_z/k_BT)}
{1+\exp((E_f-eV_a-E_z)/k_BT)}\right],
\end{eqnarray}
where $e$ is the elementary charge, $m^* = 0.145 m_0$ is the effective mass of
the electrons with $m_0$ denoting the free electron mass, $\hbar$ labels the reduced Planck constant,
$T_\sigma$ denotes the spin-dependent transmission function, $E_z$
is the longitudinal energy of the electrons, $E_f$ denotes the Fermi
energy of the left lead, and $V_a$ is the voltage applied to the
right lead.

%
%

\begin{figure}
 \centerline{\psfig{file=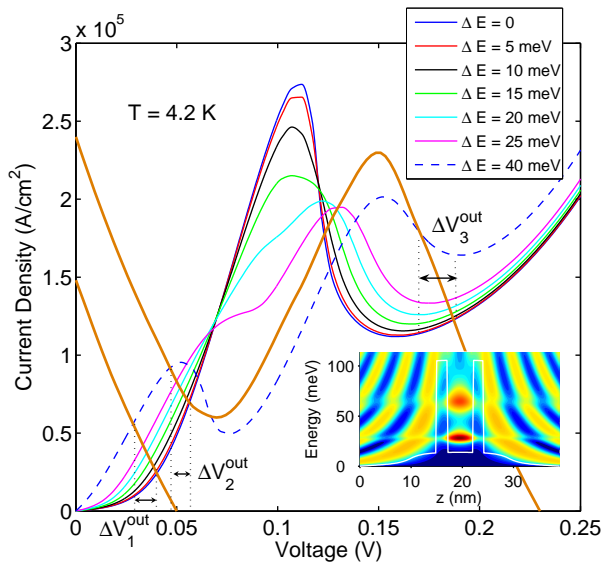,width=1\linewidth}}

\caption{(Color online) Current-Voltage characteristics of the
magnetic RTDs at the temperature of $T= 4.2$ K for different Zeeman
energy splittings $\Delta E$. The thick solid lines indicate the
mirrored IV-curve of the load RTD for low and high input voltages.
The inset shows a contour plot of the local density of states versus
energy and growth direction $z$ at zero bias and $\Delta E = 40$
meV. The solid line in the inset indicates the conduction band
profile. The two spin resonances are visible by the large density
(dark) in the well.} \label{fig:IVs}
\end{figure}

At low temperatures (we set $T = 4.2$ K as in experiment
\cite{SlGoSl03}) the Zeeman splitting is of the order of $10$ meV
for practical magnetic fields of 1-2 T and Mn concentrations of
about 8\%. The obtained IV-characteristics for the m-RTD for
different Zeeman splittings are displayed in Fig. \ref{fig:IVs}. The
inset of Fig. \ref{fig:IVs} shows a contour plot of the local
density of states at zero bias for $\Delta E = 40$ meV, with the
solid line indicating the self-consistent conduction band profile.
The upward band bending at equilibrium is caused by the undoped
buffer layers. The local density of states clearly demonstrates the
Zeeman splitting of the quasibound energy state in the quantum well.
Since we consider a positive applied magnetic field the lower and
upper energy states correspond to spin down and spin up,
respectively. For high magnetic fields the Zeeman splitting becomes
also observable in the IV-characteristics leading to two separated
current peaks. However, most important with respect to the
functionality of the proposed DMR-device is the fact that the peak
current is appreciably decreased by an applied external magnetic
field. The mirrored IV-characteristics of the nonmagnetic load
device for low (monostable regime) and high (bistable regime) input
voltages, respectively, are indicated by the thick solid lines. As
illustrated in Fig. \ref{fig:IVs} for fixed low and high input
voltage, the output voltage is restricted to three different
intervals $\Delta V_i^{\mathrm{out}}, i=1,2,3$. The high voltages
interval $\Delta V_3^{\mathrm{out}}$ are considerably separated from
the low voltages intervals $\Delta V_1^{\mathrm{out}}$ and $\Delta
V_2^{\mathrm{out}}$ which allows a digital interpretation of the
output voltage. The magnetic field dependence of m-RTDs has been
already seen experimentally in Ref. \cite{SlGoSl03}.

\begin{figure}
 \centerline{\psfig{file=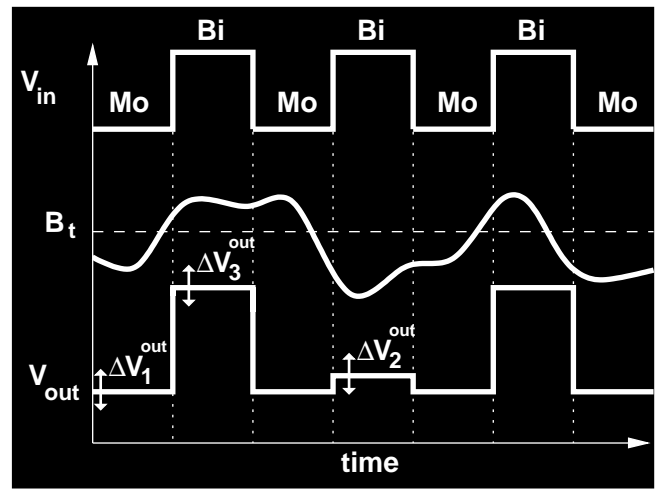,width=1\linewidth}}

\caption{Scheme of the operation principle of the DMR-MOBILE. The
vertical dotted lines indicate the points of time when a
mono-to-bistable transition is performed. The output voltages
fluctuate within the intervals $\Delta V_i^{\mathrm{out}}, i =
1,2,3$.} \label{fig:princip}
\end{figure}

To utilize the influence of the magnetic field on the driver peak
current, we suggest the following operation principle of our
DMR-MOBILE, which is schematically visualized in Fig.
\ref{fig:princip}.
%
%

Let us assume that the two RTDs of the MOBILE are of such kind that
the driver peak current is higher than the load peak current at zero
magnetic field. By applying an external magnetic field the driver
peak current is reduced and at some threshold value $B_t$ it becomes
smaller than the load peak current. This threshold is determined by
the value of the field in which the two peak currents are equal. The
magnitude of $B_t$ can be conveniently controlled by $V_G$
\cite{MaMi93}. The circuit is assumed to be driven by an oscillating
input voltage which continually performs a transition between the
mono- and bistable regime. For simplicity we consider a rectangular
input signal in Fig \ref{fig:princip}. If the magnetic field is
higher (lower) than the threshold field $B_t$ at the moment of the
transition from the monostable to the bistable regime, the output
voltage is high (low), which leads to a direct digital conversion of
the magnetic signal.

The question of how fast this conversion operation can be performed is 
closely connected to the subtle question of error rates in MOBILEs. 
The dynamical response of a m-RTD on a sudden change in the external
magnetic field is similar to the response on an abrupt change of the applied voltage
in nonmagnetic RTDs. Both operations lead effectively to a redistribution of the quasibound states 
in the quantum well, which happens on a time scale of the order of 100's of fs. However, the switching 
transition time of RTDs is dominated by the RC-time constant, which limits the 
switching performance to a few ps \cite{DiOezRo89}. 
These considerations suggest that the error rates in magnetic and conventional MOBILEs 
are comparable. In particular, parasitic capacitancies are a possible source for an erroneous mono-to-bistable
transition, since they can influence the peak
current values \cite{Waho}. Experiments show that the circuit randomly outputs high and low
in the transition region \cite{Maezawa}. These 
fluctuations might be in part due to external electric noise. 
Transient analysis of the conventional MOBILE structure 
based on equivalent circuit models \cite{Ma95} reveals that an error-free transition 
with clock rise times on the order of the RC-time of the RTD is possible if
the output capacitance $C_{\mathrm{out}}$ is reduced appropriately. A rough approximation  
yields $C_{\mathrm{out}} < (k-1) C_\mathrm{RTD}$ \cite{Ma95} where $k$ is the ratio of the 
load to driver peak current and $C_\mathrm{RTD}$ is an average capacitance of the RTD. 
Recently, an ultrahigh frequency operation of 
MOBILEs up to 100 GHz has been demonstrated by employing a symmetric clock configuration \cite{MaSuKi06}.
Hence, the proposed DMR-MOBILE scheme might be potentially used, e.g, as a very fast read head in
conventional hard disks.

To summarize, we propose a DMR-MOBILE which converts analog magnetic
signals directly into digital electrical information. The device is
essentially a MOBILE, where the driver device is replaced by a m-RTD
made of a DMS-material. How high $\Delta E$ must there be for
practical operations? One expects (and our simulations confirm) that $\Delta E$ should be
of the order of $k_BT$. With the current materials this limits the operation of the DMR-MOBILE to 
temperatures lower than perhaps 150 K. 

This work has been supported by the Deutsche Forschungsgesellschaft SFB 689. The authors thank
K. Maezawa and T. Waho for valuable discussions.



\end{document}